# Evolutionary genomics of transposable elements in *Saccharomyces cerevisiae.*

Martin Carr*[1], Douda Bensasson[2] and Casey M. Bergman[2]

[1] School of Applied Sciences, University of Huddersfield, Queensgate, Huddersfield, HD1 3DH

[2] Faculty of Life Sciences, University of Manchester, Oxford Road, Manchester, M13 9PT

<u>*Author for correspondence:</u>

School of Applied Sciences

University of Huddersfield

Queensgate

Huddersfield

HD1 3DH

United Kingdom

Telephone: +44 484-471676

email: MartCarr74@gmail.com

Version: 31 August 2012

# Abstract

*Saccharomyces cerevisiae* is one of the premier model systems for studying the genomics and evolution of transposable elements. The availability of the *S. cerevisiae* genome led to unprecedented insights into its five known transposable element families (the LTR retrotransposons *Ty1-Ty5*) in the years shortly after its completion. However, subsequent advances in bioinformatics tools for analysing transposable elements and the recent availability of genome sequences for multiple strains and species of yeast motivates new investigations into *Ty* evolution in *S. cerevisiae*.

Here we provide a comprehensive phylogenetic and population genetic analysis of all *Ty* families in *S. cerevisiae* based on a systematic re-annotation of *Ty* elements in the S288c reference genome. We show that previous annotation efforts have underestimated the total copy number of *Ty* elements for all known families. In addition, we identify a new family of *Ty3*-like elements related to the *S. paradoxus Ty3p* which is composed entirely of degenerate solo LTRs. Phylogenetic analyses of LTR sequences identified three families with short-branch, recently active clades nested among long branch, inactive insertions (*Ty1*, *Ty3*, *Ty4*), one family with essentially all recently active elements (*Ty2*) and two families with only inactive elements (*Ty3p* and *Ty5*). Population genomic data from 38 additional strains of *S. cerevisiae* show that elements present in active clades are predominantly polymorphic, whereas most of the inactive elements are fixed. Finally, we use comparative genomic data to provide evidence that the *Ty2* and *Ty3p* families have arisen in the *S. cerevisiae* genome by horizontal transfer.

Our results demonstrate that the genome of a single individual contains important information about the state of TE population dynamics within a species and suggest that horizontal transfer may play an important role in shaping the genomic diversity of transposable elements in unicellular eukaryotes.

# Introduction

Transposable elements (TEs) have been shown to be a component of most, although not all, studied eukaryotic genomes [1]. Empirical and theoretical work from a broad range of host organisms suggests that TE insertions are generally deleterious and that natural selection acts to suppress proliferation in host populations [2,3,4,5,6,7]. Classically, it has been thought that a balance between increases in copy number by transposition and selection against deleterious insertions govern the dynamics of TE evolution [8]. However, more recent work has shown that some proportion of TE insertions may also evolve under positive selection (e.g. [9,10,11,12,13]), and that TE families undergo horizontal transfer at a higher rate than previously anticipated (e.g. [14,15,16,17]). In the genomics era, many unresolved questions about the factors that control TE evolution can now be addressed with the growing volumes of available sequence data.

The first eukaryote to have its genome completely sequenced was the yeast *Saccharomyces cerevisiae* [18]. This landmark achievement allowed unprecedented insights into the impact that TEs have on genome structure and evolution. The 12.2 Mb yeast genome is known to harbour five families of TEs, *Ty1* through *Ty5*, all of which are long terminal repeat (LTR) retrotransposons [19]. *Ty1* and *Ty2* have been studied intensively and are known to be active families [20] that together make up ~75% of the *Ty* insertions in the reference genome [21,22]. These two closely-related families can be differentiated by sequence divergence in their *gag* open-reading frame (ORF) and a hypervariable region of the *pol* ORF [22]. In contrast, the only fixed difference between their LTR sequences is a one base pair deletion present in all copies of *Ty2* [22]. The high level of nucleotide identity between *Ty1* and *Ty2* LTRs permits interfamily recombination, and elements with hybrid *Ty1/2* LTRs are present in the *S. cerevisiae* genome [23]. The smaller families *Ty3-Ty5* have been characterised to a more limited extent: *Ty3* is thought to be an active family [24] with full-length elements in the genome; full-length copies of *Ty4* exist in the genome but transposition of this family has not been observed [25]; and there are no longer any functionally-intact copies of *Ty5* in the yeast genome [19].

The abundance of *Ty* elements varies considerably between populations of *S. cerevisiae*, with lab strains tending to possess greater abundance than wild populations [26,27]. The relatively low copy numbers observed in wild populations, despite ongoing transposition, is consistent with the proliferation of elements being held in check by natural selection. Work by Blanc *et al.* [28] has suggested that the presence of a single *Ty* element insert results in a <2% loss of fitness. Not all *Ty* element insertions result in deleterious mutations, however, with some *Ty* insertions being proposed to play adaptive roles in the evolution of gene regulation [29]. As with other eukaryotes, the evolutionary and genomic forces that control TE abundance and variation in yeast remain a matter of debate, necessitating detailed investigation of the copy number, allele frequency and sequence variation of *Ty* elements within and among *S. cerevisiae* genomes.

Kim *et al.* [22] conducted the first systematic survey of *Ty* insertions in the *S. cerevisiae* reference genome, determining both the copy number and genomic distribution of all five known *Ty* families and performing a phylogenetic analysis of *Ty1* and *Ty2* LTR sequences. Jordan and McDonald [30] used the element copies identified by Kim *et al.* [22] to show that all five *Ty* families exhibit low levels of sequence variation in their ORFs and appear to be evolving under purifying selection. Comparative analysis involving other species of *Saccharomyces sensu stricto* clade [31,32,33,34] provides evidence that *Ty* families have been inherited both vertically and horizontally within the *Saccharomyces* genus, as well as undergoing occasional loss from their host's genome. These insights into the evolution of *Ty* families in *Saccharomyces* species have been based primarily on single reference genomes. Genomic analysis of *Ty* elements in multiple individuals within species has, until very recently, required sophisticated microarray techniques to be devised [35,36,37]. However, with the advent of high-throughput re-sequencing approaches, it is now possible to add a population genomic perspective to the evolution of *Ty* elements in yeast species [27,38,39,40,41,42].

Here we investigate the evolutionary genomics of all *Ty* families in *S. cerevisiae* using a combination of bioinformatic, phylogenetic and population genomic approaches. We first provide a re-annotation of the *S. cerevisiae* reference genome, showing a considerably larger copy number than that originally reported by Kim *et al.* [22]. The majority of newly discovered *Ty* insertions are small, degenerate fragments that provide important data for inferring the evolutionary history of these families. In addition to finding new insertions of all five known *Ty* families, we also find insertions for a previously unannotated family, *Ty3p*. We then conduct phylogenetic and molecular evolutionary analyses on LTR sequences of all *Ty* families, which provide insight into the transpositional activity and evolutionary history of these families. For all *Ty* copies present in the reference genome, we also determine their presence or absence in a large sample of genomes from the *Saccharomyces* Genome Resequencing Project (SGRP) [27] and map this population genomic data onto *Ty* element phylogenies. In general, we find strong concordance between inferences of insertion history based on phylogenetic analysis with those based on population genomic evidence, but also identify a small subset of young *Ty* insertions at high frequency that may be under the influence of natural selection. Finally, we investigate the origins of *Ty2* and *Ty3p* in *S. cerevisiae* by phylogenetic analysis of the *Ty1/Ty2* and *Ty3/Ty3p* super-families in the *Saccharomyces sensu stricto* clade, and provide evidence that both *Ty2* and *Ty3p* in *S. cerevisiae* arose by horizontal transfer at different times in the past. Together our results demonstrate that the reference genome of a single individual contains important information about the state of TE population dynamics within a species, but that full insight into TE evolutionary genomics requires complete genome sequences from multiple individuals and species.

## Materials and Methods

*Re-annotation of Ty elements in* S. cerevisiae

The June 2008 (sacCer2) version of the *S. cerevisiae* genome was downloaded from the UCSC Genome database and screened for TE sequences using RepeatMasker open-3.1.6 using a custom yeast TE library that includes *Ty1-Ty5* from *S. cerevisiae* as well as *Ty3p* from *S. paradoxus* (File S1). Our custom TE library was constructed by exporting "saccharomyces" from the 20061006 RepeatMasker library, removing the redundant TY sequence, replacing TY4 with the first LTR element (positions 262-632) and internal sequence (positions 633-6116) from GenBank accession X67284 [43], adding the *Ty3p* LTR (Genbank: AY198187) and internal sequences (positions 306-4980 of Genbank: AY198186) [32], and renaming fasta headers for consistency. The decision to include *Ty3p* in our library was based on preliminary analyses that revealed additional *Ty3*-like sequences in the *S. cerevisiae* genome that were not fully identified using the *Ty3* element as a query.

RepeatMasker output was then processed with REANNOTATE version 26.11.2007 (options: -g -f -d 10000) [44] in order to join LTRs to internal sequences and defragment nested and degenerate insertions (see Figure 1 for an example screenshot of the REANNOTATE output). LTR sequences from *Ty1* and *Ty2* were treated as equivalent in the defragmentation process. We then compared individual *Ty* copies from the REANNOTATE output to *Ty* annotations in sacCer2 taken from SGD and a small number were manually edited to correct any obvious errors. In total, 10 *Ty* annotations from the REANNOTATE output were manually split or joined to correct defragmentation problems, and coordinate spans of an additional 19 REannotate fragments were updated to remove redundant overlaps among features.

LTR sequences of *Ty1* and *Ty2* are distinguished by a single base pair deletion in *Ty2* elements [22]. We labelled LTRs identified by REANNOTATE from either of these families as *Ty2* when this deletion could clearly be identified. We also re-evaluated the identity of all *Ty1* and *Ty2* LTRs after a preliminary phylogenetic analysis of LTRs from both families. All LTRs nested within the *Ty2* clade, including those that had secondary deletions spanning the diagnostic one bp *Ty2* deletion, were then relabelled as *Ty2* inserts. In total, 85 LTR fragments were relabelled from *Ty1* to *Ty2* or *vice versa*. The final dataset of curated *Ty* annotations can be found in File S2 and the RepeatMasker/REANNOTATE fragments supporting these annotations can be found in File S3.

*Phylogenetic analysis of paralogous LTR sequences*

Multiple alignments of paralogous LTRs were created individually for all five families (*Ty1*, *Ty2*, *Ty3+Ty3p*, *Ty4* and *Ty5*) using MUSCLE 3.7 [45]. Phylogenetic trees were generated from each alignment using maximum likelihood and Bayesian inference analyses. The maximum likelihood analyses were performed using RAxML 7.2.6 [46] on raxmlGUI 0.9.3. The analyses were initiated with 100 parsimony trees, created using the recommended GTRCAT model incorporating 25 site-rate categories, and bootstrapped with 1,000 replicates. Parameters for each analysis were determined by RAxML. Bayesian phylogenies were created using MrBayes 3.1.2 [47]. MrBayes analyses were run using a GTR+I+Γ model and a four-category gamma distribution. Searches consisted of two parallel chain sets run at default temperatures, with a sample frequency of 10, until they had reached convergence (i.e. average standard deviation of split frequencies equal to 0.01) or had ran for 5,000,000 generations. Runs consisted of a minimum of 500,000 generations. The first 25% of sampled trees were discarded as burnin before calculating posterior probabilities.

Phylogenies of the *Ty3/Ty3p* and *Ty1/Ty2* super-families were created using LTR sequences from multiple *Saccharomyces sensu stricto* species. Sequence similarity searches were performed on the genomes of *S. bayanus*, *S. castellii*, *S. kluyveri*, *S. kudriavzevii*, *S. mikatae*, *S. paradoxus*, *S. servazii* and *S. unisporus*. LTR sequences were extracted using WU-BLAST2 at SGD using the 3' LTRs of u11 (*Ty3*) and u23 (*Ty2*) as the query sequences. With the exception of the highly degenerate *Ty3p* LTR sequences in *S. cerevisiae*, hits were filtered to a minimum length of 280 bp and an E-value of $1 \times 10^{-10}$ for inclusion in these alignments. For each super-family, LTR sequences from each species were first aligned in MUSCLE and preliminary phylogenies were created from the unedited alignments using RAxML with 100 rapid bootstrap replicates using the GTRCAT model. These preliminary phylogenies were used to identify short branch (presumably recently active) sequences that spanned the diversity of active elements from each species for inclusion in the multi-species alignment. The final multispecies datasets were aligned using MUSCLE and maximum likelihood phylogenies were created following the same protocol as used for the individual *S. cerevisiae Ty* families. MrBayes analyses were also performed, running for 5,000,000 generations, a sampling frequency of 100 and with a burnin of 12,500, using the same protocol as for the individual *Ty* families. Nexus files for the *Ty3/Ty3p* and *Ty1/Ty2* super-family alignments can be found in File S4 and S5, respectively.

For the *Ty1/Ty2* superfamily the total number of hits from *S. mikatae* was 145, of which 44 were excluded on the basis of their long-branch length. LTR sequences of full-length *Ty1* elements from *S. cerevisiae* (excluding the recombinant *Ty1/Ty2* hybrid elements) were added to the dataset, as were LTR sequences of *Ty2* from *S. cerevisiae* (excluding the long branch elements s304, st39, st3, st97, st98 and st107). For the *Ty3* super-family, the *Ty3p* LTRs from *S. cerevisiae* were restricted to the seven longest sequences (which had a minimum length of 175 bp).

*Molecular evolutionary analyses of paralogous LTR sequences*

Phylogenetic trees were inspected and used to classify individual elements in the reference genome manually as being derived from 'active' or 'inactive' clades. Short-branch LTRs from clades with full-length elements were classified as active irrespective of whether the insert was itself still capable of transposition (e.g. a short-branch solo LTR from a recent intra-element recombination event would be classified as being in an active clade). LTRs present on long branches were classified as being in inactive clades. Separate alignments were created for all, inactive and active paralogous LTR sequences using MUSCLE, with only the 5' LTR included from full-length elements containing both LTRs. An archive of all the paralogous LTR alignments used for molecular evolutionary analysis can be found in File S6.

Nucleotide diversity between paralogous LTRs in the three datasets (all, inactive and active) for each family was determined in DnaSP 5.1 [48] using $\pi$ [49] and $\theta$ [50]. Tajima's *D* [51] was calculated from the alignments of active elements to detect departures from expected patterns of nucleotide diversity under a standard neutral model of sequence evolution with families that have a constant rate of insertion that are at copy number and coalescent equilibrium [52,53,54].

Recombination events between recently active elements were identified with two datasets using Hudson and Kaplan's $R_M$ [55]. First, the minimum number of recombination events ($R_M$) was determined for all full-length elements for *Ty1*, *Ty2* and *Ty4*. Second, we determined $R_M$ for all LTR sequences of active elements for *Ty1*, *Ty2*, *Ty3* and *Ty4*. This second dataset was constructed to increase the number of paralogous elements included in this analysis, albeit at the expense of sequence length. Because of limited data, *Ty3, Ty3p* and *Ty5* were excluded from analyses of recombination in full-length elements, and *Ty3p* and *Ty5* were excluded from analyses of recombination among recently active LTRs.

*Population genomic analyses of orthologous Ty insertions*

*S. cerevisiae* resequencing data was downloaded from the SGRP ftp site: ftp://ftp.sanger.ac.uk/pub/dmc/yeast/latest/misc.tgz (last modified on 20/9/08). Coordinates of re-annotated *Ty* elements in the sacCer2 reference genome were transferred to the coordinate system of the SGRP reference genome from the by extracting the entire insertion including ±200 bp of flanking sequence from the sacCer2 genome and searching for an exact match in the SGRP reference genome. SGRP based coordinates were then used to extract multiple alignments of sequenced nucleotides prior to imputation using alicat.pl (option: q40). The *S. cerevisiae* strains involved were: 273614N, 322134S, 378604X, BC187, DBVPG1106, DBVPG1373, DBVPG1788, DBVPG1853, DBVPG6040, DBVPG6044, DBVPG6765, K11, L_1374, L_1528, NCYC110, NCYC361, RM11_1A, S288c, SK1, UWOPS03_461_4, UWOPS05_217_3, UWOPS05_227_2, UWOPS83_787_3, UWOPS87_2421, W303, Y12, Y55, Y9, YIIc17_E5, YJM789, YJM975, YJM978, YJM981, YPS128, YPS606, YS2, YS4, YS9. Resulting multiple alignments were viewed in Seaview 3.2 [56] and each *Ty* insertion was manually classified as present, absent or missing data for all 38 strains of *S. cerevisiae*. Insertion sites were then classified as fixed (scored as present or missing data in all genomes), polymorphic (scored as both present and absent in genomes with data), or S288c-only (found only in the S288c reference genome). Alignments of population genomic data in the SGRP strains for each *Ty* insertion can be found in File S7.

# Results

*Re-evaluation of TE content and copy number in the S288c reference genome*

We reannotated *Ty* elements in the reference *S. cerevisiae* genome and found a total of 483 *Ty* element insertions (Table 1; Files S2 and S3), which is ~46% higher than the estimate of 331 by Kim *et al.* [22]. Increased copy number is observed for all five previously annotated *Ty* families, but is greatest for the most abundant family *Ty1* (Table 1). Our reannotation uncovered 406,829 bp of DNA (3.35%) in the *S. cerevisiae* reference genome that was of recognisable *Ty* element origin. This value is only ~30 kb greater than the estimate of Kim *et al.* [22], indicating that the majority of previously unannotated *Ty* insertions we identify are small fragments. The majority of the 152 new insertions identified here are degenerate solo LTRs, reinforcing the findings of Kim *et al.* [22] that full-length elements only make up a small proportion of *Ty* insertions in *S. cerevisiae*.

Fifteen of the newly annotated inserts were identified by including *Ty3p* (a *Ty3* family from *S. paradoxus* [32]) as a query. Evidence for *Ty3p* sequences in the *S. cerevisiae* genome has not previously been reported, although a solo LTR (YNLWsigma2) previously annotated as a *Ty3* insertion by Kim *et al.* [22] on chromosome XIV corresponds to a copy of *Ty3p*, s349, found here. *Ty3p* insertions are present only as solo LTRs in *S. cerevisiae* (Table 1 and Figure 5) and thus, like *Ty5*, this lineage is comprised entirely of non-functional insertions in *S. cerevisiae*. We do not believe that the *Ty3p* sequences we detect in *S. cerevisiae* are simply ancient copies from the *Ty3* lineage, since a phylogeny of the *Ty3* super-family (Figure 2) across the *Saccharomyces* genus showed that *Ty3* and *Ty3p* are distinct lineages that are separated by strongly supported branches (maximum likelihood bootstrap percentage, mlBP ≥70%, Bayesian inference posterior probability, biPP≥0.95). The *S. cerevisiae Ty3* elements form a sister-group with the *Ty3* sequences of *S. mikatae*, while the highly degenerate *S. cerevisiae Ty3p* insertions a found in a well-supported clade with the *Ty3p* sequences of *S. paradoxus* and *S. kudriavzevii*. Furthermore, both *Ty3* and *Ty3p* lineages are present and recently active in *S. kudriavzevii* suggesting that these are indeed distinct lineages. These results are consistent with the origin of the *Ty3p* family in *S. cerevisiae* by horizontal transfer (see below), and justify treatment of the *S. cerevisiae Ty3p* as a distinct family in the following analyses.

We identified only one additional full-length element, a *Ty1* element (u19, on chromosome XII) relative to the SGD annotation of *Ty* elements on the sacCer2 genome. Examination of this region of Chromosome XII reveals a 5,926 bp element that possesses 338 bp LTRs and encodes putative TYA (Gag) and TYB (Pol) proteins. This insertion was erroneously annotated in the sacCer2 SGD annotation as two solo LTRs (YLRCdelta2 and YLRCdelta3) [22].

Kim *et al.* [22] classified all *Ty* insertions either as full-length or solo LTRs. Our re-annotation also identified a new class of five "truncated" elements generated through deletions of genomic DNA rather than LTR-LTR recombination. Three of the truncated elements (t41, t120 and t194) were identified previously as solo LTRs [22] while the other two truncated elements (t219 and t220), which both lack any LTR sequence, have not been identified previously.

*Molecular evolution of paralogous Ty LTR sequences*

Variation among paralogous sequences of *Ty* elements in the *S. cerevisiae* reference genome allows indirect observations to be made of element activity using both phylogenetic and molecular evolutionary approaches. Recently transposed elements are unlikely to have accumulated a large number of mutations that differentiate them from their parental element; therefore recently active TEs are expected to exhibit short branches in phylogenetic trees generated from paralogous LTRs in a single genome. Likewise, active TE families that have undergone a recent increase in copy number will harbour an excess of sequence differences between paralogous copies that are at low frequency in the sample, resulting in a negative value of the Tajima's *D* statistic [53,54]. This pattern is analogous to the signature of recent population growth in population genetic analyses of an orthologous loci [51]. Finally, patterns of molecular variation that arise from genetic exchange can be used to detect gene conversion or recombination between paralogous *Ty* sequences in the past. Results for these three approaches are presented below to infer aspects of the evolutionary history for each *Ty* family.

**Ty1**

*Ty1* has a much higher copy number (n=313) than all other *Ty* families present in the *S. cerevisiae* genome (Table 1; see also [22]). Long branch, presumably inactive, insertions make up the majority (~87%, n=272; Table 2) of *Ty1* insertions; these are exclusively solo LTRs (Figure 3). Short branch, presumably active *Ty1* elements cluster together on the phylogeny, suggesting that all such elements have a recent common ancestry. There are a total of 41 *Ty1* elements that appear to have inserted recently into the reference genome (Table 2, Figure 3). Eight of these young elements have subsequently undergone intra-element LTR-LTR recombination to become solo LTRs.

The phylogeny also supports subdivision of the active *Ty1* lineages into three previously proposed subfamilies [22,23] that we refer to as "canonical" *Ty1*, *Ty1'* and *Ty1/Ty2* (Figure 3). The majority of *Ty1* insertions are from the canonical, presumably ancestral, subfamily. In the clade containing the active canonical class, there are fourteen full-length elements and one solo LTR. The *Ty1'* subfamily [22] is characterised by its highly divergent *gag* ORF. In addition to the three full-length elements described by Kim *et al.* [22], we identify a further five solo LTRs in the *Ty1'* subfamily. The *Ty1* phylogeny also supports a clade of 18 recombinant *Ty1/2* elements [23], which fall into two distinct groups. The maximum likelihood tree places five long-branch solo LTRs within the *Ty1/2* clade (shown as collapsed red clades in Figure 3), however their positions are not strongly supported and they are not recovered in this position in the Bayesian tree (data not shown). Fourteen of the *Ty1/2* elements have previously been described and the approximate sites of recombination mapped [23,57]. The newly identified hybrid elements appear to show the same recombination breakpoints (File S8).

LTR sequences of the active *Ty1* elements harbour greater nucleotide diversity in comparison to the LTRs of active elements from other families (Table 2). Tajima's *D* for the active *Ty1* elements is positive, rather than negative (Table 2). It is unlikely that this result is due to a recent contraction in the number of elements within the genome; the phylogeny strongly suggests that *Ty1* is currently transposing in *S. cerevisiae*, a result confirmed in numerous studies (e.g. [58,59]). A more probable explanation is the presence of multiple active *Ty1* subfamilies (see above, Figure 3). This phylogenetic sub-structure (like population subdivision) will result in intermediate frequency polymorphisms in the total population of active elements, leading to a positive value of $D$ among paralogous *Ty1* elements within the reference genome. Separate alignments of only canonical, *Ty1'*, or *Ty1/2* elements however all exhibit a negative value for Tajima's $D$ (data not shown), consistent with their recent transposition.

Patterns of sequence diversity provide evidence for recombination between the full-length *Ty1* copies that have integrated into the genome, with a minimum of 41 recombination events occurring between the 32 full-length elements. These results are consistent with recombination among paralogous sequences occurring at low frequency [60].

**Ty2**

The total copy number of *Ty2* (n=46) is similar to those of *Ty3* (n=45) and *Ty4* (n=49), but substantially lower than *Ty1* (Table 1). The *Ty2* phylogeny is poorly resolved under both maximum likelihood and Bayesian inference methods (Figure 4). Unlike *Ty1*, *Ty3* and *Ty4* there is no clear separation of long-branch, inactive elements and short branch, putatively active elements. There are nine long-branch solo LTRs in the phylogeny, which are presumably older inserts, however they do not cluster together. The remaining 35 inserts are present on short branches and appear to have recently integrated. Intra-element LTR-LTR recombination or truncation has converted 23 of the 35 recently integrated copies to solo LTRs.

Consistent with a poorly resolved phylogeny, *Ty2* elements have diverged less in comparison to all other families in the *S. cerevisiae* genome. Recent common ancestry of most *Ty2* copies seems to be the most likely explanation of the poorly resolved phylogeny and relatively low nucleotide diversity in comparison to the other TE families in the genome (see below). A negative value of Tajima's $D$ is consistent with the recent arrival of *Ty2* elements in the *S. cerevisiae* genome (Table 2).

In spite of their recent origin, there is evidence that paralogous copies of *Ty2* are undergoing recombination, similar to *Ty1*. A minimum of 19 recombination events can be detected between the 13 full-length copies of *Ty2* and recently integrated *Ty2* LTR sequences also show evidence for three recombination events among paralogous copies.

**Ty3**

The 45 copies of *Ty3* exhibit lower levels of nucleotide diversity than most other families, with only copies of *Ty2* showing less variation (Table 2). The *Ty3* phylogeny exhibits long-branch, older inserts as well as a single clade containing putatively active elements (Figure 5). Nested among the short branch, younger inserts are two clades of long branch LTRs. The placements of the two long branch clades have no support (mlBP <50%, biPP<0.70), are not supported by the Bayesian phylogeny, and their placement in the tree are most likely to be phylogenetic artefacts.

The active clade of *Ty3* (excluding the long branch LTRs) is composed of 24 inserts and shows the lowest nucleotide diversity of all the active *Ty* lineages. Tajima's *D* calculated from the alignment of LTRs from this clade is negative, consistent with this group of elements having undergone a recent expansion (Table 2). Although this clade has recently been active there are only two full-length, and therefore potentially transpositionally competent, elements in the reference genome. The remainder of the inserts within this clade are solo LTRs, suggesting a high rate of intra-element LTR-LTR recombination for this family The proportion of solo LTRs seen in the active *Ty3* lineage is not significantly different to that seen in the active lineage of *Ty4* elements (Fisher's exact test; $P$=0.36), however it is significantly higher than both *Ty1* ($P$<0.0001) and *Ty2* ($P$<0.0001).

As there are only two full-length copies of *Ty3* in the reference genome, it was not possible to detect any possible recombination events between full-length paralogues. Nucleotide sequences of the recently inserted copies of *Ty3* do not show any evidence of recombination between the LTR sequences at paralogous sites.

**Ty3p**

The 15 copies of the new *Ty3p* family we report here are only present as highly degenerate solo LTRs (Table 1). This family also harbours the highest level of nucleotide diversity among paralogous copies in the reference genome (Table 2). In addition, no copies of *Ty3p* are present on short branches in the combined *Ty3* and *Ty3p* phylogeny (Figure 5), highlighting a lack of recent transposition for this family. Together, these features indicate that *Ty3p* is an ancient component of the genome that lost the ability to transpose in *S. cerevisiae* long ago. Because there are no LTRs from full-length elements for this family, we did not perform recombination analyses for *Ty3p*.

**Ty4**

The phylogeny of the 49 Ty4 insertions shows a single, well-supported (72% mlBP, 0.77 biPP) recently active lineage of 16 insertions with the remainder of the 33 LTR sequences being present on long, poorly supported branches (Figure 6). Only three of the insertions from the recently active clade are present as full-length copies in the reference genome, since the remainder have undergone LTR-LTR recombination. Tajima's *D* is negative for the active clade of *Ty4* (Table 2), consistent with recent transposition for this group of elements. Hudson and Kaplan's $R_M$ failed to find any evidence of recombination between the full-length copies of *Ty4* or between the active LTR sequences of this family.

**Ty5**

Phylogenetic analysis of the 15 *Ty5* insertions revealed the absence of a putatively active clade for this family, suggesting that there has been very little, if any, transposition of this family in the recent evolutionary history of *S. cerevisiae* (Figure 7). This result supports previous findings [61] and is reinforced by the fact that the only full-length copy of *Ty5* in the reference genome harbours internal deletions and the LTRs share 91% identity [19]. There are two identical *Ty5* solo LTRs on chromosomes IV and X (s49 and s261) that give rise to short branches in the phylogeny; however examination of their flanking DNA shows that they have arisen due to an inter-chromosomal segmental duplication rather than transposition, as has been observed previously for *Ty1* [22]. The same genomic duplication event also generated additional copies of the *Ty4* solo LTRs that flank the s49 and s261 *Ty5* fragments, since the *Ty4* inserts s50 and s260 are also present in the duplicated region. Because there are no LTRs from full-length elements for this family, we did not perform recombination analyses for *Ty5*.

*Population genomic analyses of Ty elements in 38* **S. cerevisiae** *strains*

While important inferences about the history of *Ty* element activity such as those above can be made from a single reference genome, it remains an open question whether information in reference genomes accurately represent the state of TE population dynamics within a species. To address this issue, all 483 *Ty* element insertions present in the reference genome were screened *in silico* for their presence or absence in whole genome reference-based alignments of the 37 other *S. cerevisiae* strains sequenced by the SGRP. Because we only use the actual, not imputed, sequence data from the SGRP alignments, the relevant segment of an SGRP strain can be missing due to the low-coverage nature of these shotgun sequences. The number of genomes with data available to score presence/absence of an insertion ranged from 2-37 per locus and, in total, 44% of alleles were classified as missing data across all strains and loci (see File S2). Because of the substantial degree of missing data, we did not attempt to analyze allele frequencies for each insertion and instead chose to classify *Ty* insertions as fixed, polymorphic or S288c-only. Polymorphic *Ty* insertions may be misclassified as S288c-only because of missing data in other strains, and thus both classes can be considered polymorphic.

Based on presence/absence data in the SGRP alignments, we estimate that 73.7% (356/483) of the *Ty* elements in the reference genome are fixed (scored as present in all genomes with sequence data) in the *S. cerevisiae* population (Table 1). *Ty2* shows the lowest proportion of fixed elements (23.9%), consistent with most insertions from this family having recently integrated into the *S. cerevisiae* genome. Nearly half of the *Ty3* insertions are fixed (48.9%), and the majority of elements in the other *Ty* families are fixed. Fourteen of the fifteen *Ty3p* solo LTRs are fixed within the scored genomes. Only one copy appears to be polymorphic, but this is due to a post-fixation secondary deletion in the strain RM11_1A that covers the insertion site as well as the flanking DNA. Fixation of all *Ty3p* insertions is consistent with lack of recent transposition for this family in the *S. cerevisiae* genome. Previous studies have reported a lack of *Ty5* in some strains of *S. cerevisiae* [34,61]; our *in silico* screen of the SGRP genomes however shows that *Ty5* copies are predominantly fixed, with only two segregating inserts present in the reference genome. The two polymorphic inserts are both solo LTRs and are adjacent to each other, albeit on different strands, on Chromosome V.

We mapped polymorphism or fixation states of each insert onto the LTR phylogenies, which showed that most of the long-branch, inactive inserts are fixed within the population. In contrast, almost all of the inserts in the putatively active clades are polymorphic (Figures 3, 4, 5, 6, 7). This result demonstrates that both terminal branch lengths estimated from phylogenies of paralogous LTRs and fixation status of insertions estimated from population genomic data provide consistent inferences about the history of individual *Ty* insertions in *S. cerevisiae*. The mean terminal branch lengths of fixed elements for *Ty1*, *Ty2*, *Ty3* and *Ty4* are an order of magnitude or more greater than the mean terminal branch lengths of polymorphic inserts (Table 2). The mean terminal branch length for fixed copies of *Ty1*, *Ty3* and *Ty4* are of a similar scale, suggesting that they have been components of the *S. cerevisiae* genome for a similar length of time. In contrast, fixed copies of *Ty2* show considerably shorter terminal branch lengths, which is consistent with a more recent origin of *Ty2* in *S. cerevisiae*.

***Horizontal transfer explains the existence of closely related Ty families in* S. cerevisiae**

Two contrasting models can be put forward to explain the existence of closely related *Ty* families (like *Ty1* and *Ty2*, or *Ty3* and *Ty3p*) in the *S. cerevisiae* genome [22]. The first model proposes that closely-related families are sub-lineages of an ancestral super-family that evolved by vertical transmission and TE "speciation" within *S. cerevisiae* or its ancestor. Evidence for distinct, active subfamilies of *Ty1* in *S. cerevisiae* support the plausibility of the TE speciation model (Figure 3).

The second model proposes that closely-related families are sub-lineages of the same super-family that arose in different *Saccharomyces* species and subsequently came together in *S. cerevisiae* or its ancestor by horizontal transfer [22,34]. In the case of *Ty1* and *Ty2*, support for the horizontal transfer model comes from the fact that some *Saccharomyces* species (e.g. *S. paradoxus* and *S. bayanus*) appear to have only *Ty1* but not *Ty2* elements [62], while *S. mikatae* appears to have only *Ty2* but not *Ty1* elements [63]. Furthermore, Liti *et al.* [34] noted that *Ty2* is present in *S. mikatae* but not in species more closely related to *S. cerevisiae*, and that *Ty2* from *S. cerevisiae* and *S. mikatae* share a high level of nucleotide identity. These observations led Liti *et al.* [34] to propose a horizontal transfer of *Ty2* between *S. cerevisiae* and *S. mikatae*, but based on available data these authors could not determine the direction of any putative horizontal transfer event.

The low degree of nucleotide diversity observed here in *Ty2* is consistent with a recent invasion of the *S. cerevisiae* genome by this family. Furthermore *Ty2* has a significantly lower proportion of both long-branch and fixed copies in comparison to other families, and its fixed copies exhibit a considerably shorter mean terminal branch length (Table 2). The recent origin of virtually all *Ty2* copies in *S. cerevisiae* and the absence of *Ty2* in *S. paradoxus* – other than in hybrid strains [34] – point to the direction of horizontal transfer being from *S. mikatae* to *S. cerevisiae*. To test this model, we conducted a phylogenetic analysis of active *Ty1* and *Ty2* LTR sequences from multiple *Saccharomyces sensu stricto* species (Figure 8). This analysis provides direct evidence that *Ty2* is not a subfamily of the *S. cerevisiae Ty1* family, as the *Ty2* families of *S. cerevisiae* and *S. mikatae* robustly cluster together (92% mlBP, 0.97 biPP). The phylogeny nests the monophyletic *S. cerevisiae Ty2* elements within the *S. mikatae Ty2* elements, with strong support (74% mlBP, 0.97 biPP), providing clear evidence that the direction of the horizontal transfer was from *S. mikatae* to *S. cerevisiae*. The observed phylogeny is not simply the result of LTR swapping through recombination, as protein phylogenies of *Gag* and a 300 amino acid residue region of the of *Pol* both show that the *S. cerevisiae Ty2* is more closely related to *Ty2* of *S. mikatae* than *S. cerevisiae Ty1* (data not shown).

Support for horizontal transfer as the primary explanation to understand the existence of closely related *Ty* families within a yeast species is also found in our *Ty3/Ty3p* superfamily analysis (Figure 2). However, the observation of distinct *Ty3* and *Ty3p* lineages in *S. cerevisiae* and *S. kudriavzevii* is compatible with several possible scenarios involving horizontal transfer. Under one scenario, a single *Ty3* family would have arisen in the ancestral *Saccharomyces* species with vertical transmission leading to a single *Ty3*-like family in most species except *S. cerevisiae* and *S. kudriavzevii*, which have two *Ty3*-like families due to horizontal transfer events: (i) one unsuccessful transfer of *Ty3p* from *S. paradoxus* to *S. cerevisiae* and (ii) one successful transfer of *Ty3p* from *S. paradoxus* to *S. kudriavzevii*. Alternatively, the presence of *S. cerevisiae* and *S. kudriavzevii* could be explained by speciation of the *Ty3*-like ancestor into *Ty3* and *Ty3p* along the lineage leading to *S. kudriavzevii*, an ancient horizontal transfer of the *S. kudriavzevii Ty3p* family into the ancestor of *S. cerevisiae* and *S. paradoxus*, followed by inactivation of *Ty3p* in *S. cerevisiae* and complete loss of *Ty3* in *S. paradoxus*. Several other plausible alternatives that require at least one horizontal transfer event can also be formulated. Evaluation of which scenario can best explain the data will require further systematic genomic analysis in multiple *Saccharomyces* species. We note however that, while at face value our conclusion that *Ty3* has undergone horizontal transfer seems to contradict previous claims [32], our results are in fact compatible with the data of Fingerman *et al.* [32] which only support a long-term association of *Ty3*-like lineages in the *Saccharomyces* genus, but do not actually preclude the possibility of horizontal transfer events within the genus.

# Discussion

The transposable elements of *S. cerevisiae* represent the most comprehensively investigated set of TEs studied at the genomic level within the eukaryotes. However our current understanding of evolutionary trends within *Ty* elements is predominantly based on research conducted at the end of the last century using data from a single reference genome [22,23,30,64]. Furthermore, previous studies focussed almost exclusively on a subset of families (namely *Ty1* and *Ty2*), and essentially no inferences have been drawn about the history of the smaller families (*Ty3*, *Ty3p*, *Ty4* and *Ty5*) (but see [30]). With more sophisticated bioinformatic techniques now available to study TEs in genome sequences [65], as well as population and comparative genomic data from multiple *Saccharomyces* strains and species [27,66,67], it is an opportune time to re-evaluate the evolution of *Ty* elements in *S. cerevisiae*. The analyses performed here have shown that many of the ideas on TE evolution in *S. cerevisiae* require substantial revision.

Our high-resolution re-annotation of *Ty* elements presented here reveals that the original survey of *Ty* elements in the *S. cerevisiae* genome underestimated copy number by almost 50% [22]. Kim *et al.* [22] used cutoffs that allowed the identification of all full-length elements in the reference genome, but omitted a large number of additional partial elements that we have detected with improved bioinformatic methods and a refined library of *Ty* sequences. One of the more surprising aspects of the re-analysis is the identification of a previously un-annotated family of LTR retrotransposons, *Ty3p*. In addition to improving the *S. cerevisiae* genome annotation and providing resources to shed light on the evolution of transposable elements, the additional *Ty* sequences found here have practical importance for efforts to engineer synthetic yeast chromosomes that lack destabilising sequences like *Ty* elements and tRNA genes [68].

Our results also suggest that some conclusions about the long-term co-evolutionary relationship of *Ty* elements with their host should be re-interpreted. Kim *et al.* [22] proposed that *Ty3* and *Ty4* may be recent arrivals in the *S. cerevisiae* genome, due to the low average nucleotide diversity that they observed in both families. In contrast, our phylogenetic analysis reveals long-branch, ancient lineages in these families indicating that both families are long-term inhabitants of the *S. cerevisiae* genome.

*Ty3* and *Ty4* both also exhibit short-branch clades in their phylogenies providing evidence for the recent, and probably current, activity of these families in the *S. cerevisiae* population. This inference is supported by the negative value of Tajima's *D* for the putatively active lineages of *Ty3* and *Ty4*, which implies recent growth in the number of elements for these families within the yeast genome.

Our work also resolves controversy about the origin of the *Ty2* family, which has been proposed to have evolved either from a *Ty1*-like ancestor or arisen by horizontal transfer [22,34]. Both the lack of old copies of *Ty2* in the *S. cerevisiae* genome and the phylogenetic position within the *Ty1/Ty2* super-family are consistent with the recent horizontal transfer of this family, with *S. mikatae* the most probable donor species. The ability of *Saccharomyces sensu stricto* species to form viable hybrids naturally in the wild (reviewed in [69]) provides a simple mechanism for the horizontal transfer of TEs between species without the requirement of an intermediate vector. Interspecific hybrids typically show dramatically reduced fertility (reviewed in [69]), but viable gametes can be produced by interspecific hybrids [70,71,72] that can backcross with their parental species [72]. Moreover, several studies have reported naturally-occurring introgressions between *S. cerevisiae* and *S. paradoxus* [73,74,75,76] demonstrating the successful transfer of genetic information between *Saccharomyces* species. If interspecific hybridisation is involved in the transfer of *Ty* families between *Saccharomyces* species it must be rare, since *Ty* element sequences are clearly not homogenized across yeast species [34].

The availability of genomic data from multiple *S. cerevisiae* strains means that studies on the evolution of *Ty* elements in this species are no longer limited to studies on a single reference genome. Population genomic analyses show that, despite the assumed deleterious nature of most TE insertions, a surprisingly high percentage (73.7%) of *Ty* copies are fixed in *S. cerevisiae*. These results support previous work that has shown many *Ty* LTR sequences in *S. paradoxus* are fixed in natural populations [32,77], permitting their use as putatively unconstrained sequences in molecular evolutionary studies [77,78]. The vast majority (98%) of fixed *Ty* insertions are partial elements. The dearth of fixed full-length elements may be due in part to LTR-LTR recombination being a one-way process that occurs more rapidly than the sojourn time to fixation. However it is also likely that solo LTRs are less deleterious to their hosts [7], and are therefore more likely to go to fixation by drift. Regardless of the processes that lead to fixation, the fact that most *Ty* insertions are fixed suggests that the catalogue of *Ty* insertion sites discovered using the S288c reference genome describes the core state of most *Ty* element locations across strains in this species. Moreover, since S288c has a high proportion of *Ty* sequences than other strains [27], population genomic analysis based on discovery of *Ty* elements in this reference strain are likely to survey a greater number of potential insertion sites than those based on other *S. cerevisiae* strains.

While the majority of fixed elements are old *Ty* insertions, we have identified 33 young insertions that are segregating at high frequencies ($\geq 0.50$, >15 scored genomes) in sequenced *S. cerevisiae* strains (File S2), seven of which are full-length copies and 26 are solo LTRs. These insertions may be at high frequency due to drift or hitchhiking along with advantageous host alleles, but it is also possible that a number of these elements are positively selected. The LTRs of *Ty* elements contain multiple modifiers of transcription [25,79,80], and TEs are known to be able to alter the expression of neighbouring host genes [80,81]. However, none of the high frequency *Ty* insertions identified here correspond to a *Ty1* insertion previously reported to confer an adaptive change in gene expression of the *HAP1* gene [29,82]. Nevertheless, the high-frequency, young insertions identified here present a candidate set for further investigation into potential advantageous *Ty* insertions in *S. cerevisiae*.

By combining phylogenetic inferences with population genomic data on the presence or absence of *Ty* elements, our work provides important confirmation that terminal branch lengths in phylogenetic trees provide information about the population frequency of individual insertions. Long branch inserts tend to be fixed in the *S. cerevisiae* genome, whereas short branch inserts mainly segregate at lower frequencies. This observation supports many previous studies (e.g. [7,22,83,84]) that rely on information from a single genome to make inferences about the dynamics of TE evolution in a given species. As we enter the era of widespread population-genomic resequencing, it will be of considerable interest to see how commonly this trend holds for LTR elements and other classes of TEs in the genomes and populations of other eukaryote species.

## Acknowledgements

We thank Jainy Shah for preliminary analysis of *Ty* re-annotation, Killian McSorley for preliminary analysis of *Ty4* and *Ty5* phylogenies, Vini Periera for assistance interpreting REannotate output, and Daniela Delneri for helpful discussion. We also thank Daniela Delneri and Ben Blackburne for critical comments on the manuscript.

## Author Contributions

Conceived and designed the experiments: MC, DB, CMB. Performed the experiments: MC, DB, CMB. Analyzed the data: MC, CMB. Contributed reagents/materials/analysis tools: DB, CMB. Wrote the paper: MC, CMB.

# References

1. Pritham EJ (2009) Transposable elements and factors influencing their success in eukaryotes. J Hered 100: 648-655.
2. Montgomery E, Charlesworth B, Langley CH (1987) A test for the role of natural selection in the stabilization of transposable element copy number in a population of *Drosophila melanogaster*. Genet Res 49: 31-41.
3. Biemont C, Tsitrone A, Vieira C, Hoogland C (1997) Transposable element distribution in Drosophila. Genetics 147: 1997-1999.
4. Charlesworth B, Langley CH, Sniegowski PD (1997) Transposable element distributions in *Drosophila*. Genetics 147: 1993-1995.
5. Hollister JD, Gaut BS (2007) Population and evolutionary dynamics of Helitron transposable elements in Arabidopsis thaliana. Mol Biol Evol 24: 2515-2524.
6. Song M, Boissinot S (2007) Selection against LINE-1 retrotransposons results principally from their ability to mediate ectopic recombination. Gene 390: 206-213.
7. Carr M, Nelson M, Leadbeater BS, Baldauf SL (2008) Three families of LTR retrotransposons are present in the genome of the choanoflagellate Monosiga brevicollis. Protist 159: 579-590.
8. Charlesworth B, Langley CH (1989) The population genetics of Drosophila transposable elements. Annu Rev Genet 23: 251-287.
9. Daborn PJ, Yen JL, Bogwitz MR, Le Goff G, Feil E, et al. (2002) A single p450 allele associated with insecticide resistance in Drosophila. Science 297: 2253-2256.
10. Darboux I, Charles JF, Pauchet Y, Warot S, Pauron D (2007) Transposon-mediated resistance to Bacillus sphaericus in a field-evolved population of Culex pipiens (Diptera: Culicidae). Cell Microbiol 9: 2022-2029.
11. Schlenke TA, Begun DJ (2004) Strong selective sweep associated with a transposon insertion in Drosophila simulans. Proc Natl Acad Sci U S A 101: 1626-1631.
12. Gonzalez J, Lenkov K, Lipatov M, Macpherson JM, Petrov DA (2008) High rate of recent transposable element-induced adaptation in Drosophila melanogaster. PLoS Biol 6: e251.
13. Aminetzach YT, Macpherson JM, Petrov DA (2005) Pesticide resistance via transposition-mediated adaptive gene truncation in *Drosophila*. Science 309: 764-767.
14. Bartolome C, Bello X, Maside X (2009) Widespread evidence for horizontal transfer of transposable elements across Drosophila genomes. Genome Biol 10: R22.
15. Lerat E, Burlet N, Biemont C, Vieira C (2011) Comparative analysis of transposable elements in the melanogaster subgroup sequenced genomes. Gene 473: 100-109.
16. Thomas J, Schaack S, Pritham EJ (2010) Pervasive horizontal transfer of rolling-circle transposons among animals. Genome Biol Evol 2: 656-664.
17. Gilbert C, Schaack S, Pace JK, 2nd, Brindley PJ, Feschotte C (2010) A role for host-parasite interactions in the horizontal transfer of transposons across phyla. Nature 464: 1347-1350.
18. Goffeau A, Barrell BG, Bussey H, Davis RW, Dujon B, et al. (1996) Life with 6000 genes. Science 274: 546, 563-547.
19. Voytas DF, Boeke JD (1992) Yeast retrotransposon revealed. Nature 358: 717.

20. Kingsman AJ, Gimlich RL, Clarke L, Chinault AC, Carbon J (1981) Sequence variation in dispersed repetitive sequences in Saccharomyces cerevisiae. J Mol Biol 145: 619-632.
21. Hani J, Feldmann H (1998) tRNA genes and retroelements in the yeast genome. Nucleic Acids Res 26: 689-696.
22. Kim JM, Vanguri S, Boeke JD, Gabriel A, Voytas DF (1998) Transposable elements and genome organization: a comprehensive survey of retrotransposons revealed by the complete Saccharomyces cerevisiae genome sequence. Genome Res 8: 464-478.
23. Jordan IK, McDonald JF (1998) Evidence for the role of recombination in the regulatory evolution of Saccharomyces cerevisiae Ty elements. J Mol Evol 47: 14-20.
24. Hansen LJ, Sandmeyer SB (1990) Characterization of a transpositionally active Ty3 element and identification of the Ty3 integrase protein. J Virol 64: 2599-2607.
25. Hug AM, Feldmann H (1996) Yeast retrotransposon Ty4: the majority of the rare transcripts lack a U3-R sequence. Nucleic Acids Res 24: 2338-2346.
26. Wilke CM, Maimer E, Adams J (1992) The population biology and evolutionary significance of Ty elements in Saccharomyces cerevisiae. Genetica 86: 155-173.
27. Liti G, Carter DM, Moses AM, Warringer J, Parts L, et al. (2009) Population genomics of domestic and wild yeasts. Nature 458: 337-341.
28. Blanc VM, Adams J (2004) Ty1 insertions in intergenic regions of the genome of Saccharomyces cerevisiae transcribed by RNA polymerase III have no detectable selective effect. FEMS Yeast Res 4: 487-491.
29. Fraser HB, Moses AM, Schadt EE (2010) Evidence for widespread adaptive evolution of gene expression in budding yeast. Proc Natl Acad Sci U S A 107: 2977-2982.
30. Jordan IK, McDonald JF (1999) Tempo and mode of Ty element evolution in Saccharomyces cerevisiae. Genetics 151: 1341-1351.
31. Neuveglise C, Feldmann H, Bon E, Gaillardin C, Casaregola S (2002) Genomic evolution of the long terminal repeat retrotransposons in hemiascomycetous yeasts. Genome Res 12: 930-943.
32. Fingerman EG, Dombrowski PG, Francis CA, Sniegowski PD (2003) Distribution and sequence analysis of a novel Ty3-like element in natural Saccharomyces paradoxus isolates. Yeast 20: 761-770.
33. Moore SP, Liti G, Stefanisko KM, Nyswaner KM, Chang C, et al. (2004) Analysis of a Ty1-less variant of Saccharomyces paradoxus: the gain and loss of Ty1 elements. Yeast 21: 649-660.
34. Liti G, Peruffo A, James SA, Roberts IN, Louis EJ (2005) Inferences of evolutionary relationships from a population survey of LTR-retrotransposons and telomeric-associated sequences in the Saccharomyces sensu stricto complex. Yeast 22: 177-192.
35. Wheelan SJ, Scheifele LZ, Martinez-Murillo F, Irizarry RA, Boeke JD (2006) Transposon insertion site profiling chip (TIP-chip). Proc Natl Acad Sci U S A 103: 17632-17637.
36. Gabriel A, Dapprich J, Kunkel M, Gresham D, Pratt SC, et al. (2006) Global mapping of transposon location. PLoS Genet 2: e212.

37. Carreto L, Eiriz MF, Gomes AC, Pereira PM, Schuller D, et al. (2008) Comparative genomics of wild type yeast strains unveils important genome diversity. BMC Genomics 9: 524.
38. Shibata Y, Malhotra A, Bekiranov S, Dutta A (2009) Yeast genome analysis identifies chromosomal translocation, gene conversion events and several sites of Ty element insertion. Nucleic Acids Res 37: 6454-6465.
39. Mularoni L, Zhou Y, Bowen T, Gangadharan S, Wheelan S, et al. (2012) Retrotransposon Ty1 integration targets specifically positioned asymmetric nucleosomal DNA segments in tRNA hotspots. Genome Res.
40. Qi X, Daily K, Nguyen K, Wang H, Mayhew D, et al. (2012) Retrotransposon profiling of RNA polymerase III initiation sites. Genome Research.
41. Baller JA, Gao J, Voytas DF (2011) Access to DNA establishes a secondary target site bias for the yeast retrotransposon Ty5. Proc Natl Acad Sci U S A 108: 20351-20356.
42. Baller JA, Gao J, Stamenova R, Curcio MJ, Voytas DF (2012) A nucleosomal surface defines an integration hotspot for the Saccharomyces cerevisiae Ty1 retrotransposon. Genome Res.
43. Janetzky B, Lehle L (1992) Ty4, a new retrotransposon from Saccharomyces cerevisiae, flanked by tau-elements. J Biol Chem 267: 19798-19805.
44. Pereira V (2008) Automated paleontology of repetitive DNA with REANNOTATE. BMC Genomics 9: 614.
45. Edgar RC (2004) MUSCLE: multiple sequence alignment with high accuracy and high throughput. Nucleic Acids Res 32: 1792-1797.
46. Stamatakis A, Hoover P, Rougemont J (2008) A rapid bootstrap algorithm for the RAxML Web servers. Syst Biol 57: 758-771.
47. Ronquist F, Huelsenbeck JP (2003) MrBayes 3: Bayesian phylogenetic inference under mixed models. Bioinformatics 19: 1572-1574.
48. Librado P, Rozas J (2009) DnaSP v5: a software for comprehensive analysis of DNA polymorphism data. Bioinformatics 25: 1451-1452.
49. Nei M, Li WH (1979) Mathematical model for studying genetic variation in terms of restriction endonucleases. Proc Natl Acad Sci U S A 76: 5269-5273.
50. Watterson GA (1975) On the number of segregating sites in genetical models without recombination. Theor Popul Biol 7: 256-276.
51. Tajima F (1989) Statistical method for testing the neutral mutation hypothesis by DNA polymorphism. Genetics 123: 585-595.
52. Brookfield JF (1986) A model for DNA sequence evolution within transposable element families. Genetics 112: 393-407.
53. Maside X, Bartolome C, Charlesworth B (2003) Inferences on the evolutionary history of the S-element family of Drosophila melanogaster. Mol Biol Evol 20: 1183-1187.
54. Sanchez-Gracia A, Maside X, Charlesworth B (2005) High rate of horizontal transfer of transposable elements in Drosophila. Trends Genet 21: 200-203.
55. Hudson RR, Kaplan NL (1985) Statistical properties of the number of recombination events in the history of a sample of DNA sequences. Genetics 111: 147-164.
56. Galtier N, Gouy M, Gautier C (1996) SEAVIEW and PHYLO_WIN: two graphic tools for sequence alignment and molecular phylogeny. Comput Appl Biosci 12: 543-548.
57. Jordan IK, McDonald JF (1999) Comparative genomics and evolutionary dynamics of Saccharomyces cerevisiae Ty elements. Genetica 107: 3-13.

58. Cameron JR, Loh EY, Davis RW (1979) Evidence for transposition of dispersed repetitive DNA families in yeast. Cell 16: 739-751.
59. Curcio MJ, Sanders NJ, Garfinkel DJ (1988) Transpositional competence and transcription of endogenous Ty elements in Saccharomyces cerevisiae: implications for regulation of transposition. Mol Cell Biol 8: 3571-3581.
60. Kupiec M, Petes TD (1988) Meiotic recombination between repeated transposable elements in Saccharomyces cerevisiae. Mol Cell Biol 8: 2942-2954.
61. Zou S, Wright DA, Voytas DF (1995) The Saccharomyces Ty5 retrotransposon family is associated with origins of DNA replication at the telomeres and the silent mating locus HMR. Proc Natl Acad Sci U S A 92: 920-924.
62. Naumov GI, Naumova ES, Lantto RA, Louis EJ, Korhola M (1992) Genetic homology between Saccharomyces cerevisiae and its sibling species S. paradoxus and S. bayanus: electrophoretic karyotypes. Yeast 8: 599-612.
63. Fink GR, Boeke JD, Garfinkel DJ (1986) The mechanism and consequences of retrotransposition. Trends in Genetics 2: 118-123.
64. Jordan IK, McDonald JF (1998) Interelement selection in the regulatory region of the copia retrotransposon. J Mol Evol 47: 670-676.
65. Bergman CM, Quesneville H (2007) Discovering and detecting transposable elements in genome sequences. Brief Bioinform 8: 382-392.
66. Kellis M, Patterson N, Endrizzi M, Birren B, Lander ES (2003) Sequencing and comparison of yeast species to identify genes and regulatory elements. Nature 423: 241-254.
67. Cliften P, Sudarsanam P, Desikan A, Fulton L, Fulton B, et al. (2003) Finding functional features in Saccharomyces genomes by phylogenetic footprinting. Science 301: 71-76.
68. Dymond JS, Richardson SM, Coombes CE, Babatz T, Muller H, et al. (2011) Synthetic chromosome arms function in yeast and generate phenotypic diversity by design. Nature 477: 471-476.
69. Sipiczki M (2008) Interspecies hybridization and recombination in Saccharomyces wine yeasts. FEMS Yeast Res 8: 996-1007.
70. Marinoni G, Manuel M, Petersen RF, Hvidtfeldt J, Sulo P, et al. (1999) Horizontal transfer of genetic material among Saccharomyces yeasts. J Bacteriol 181: 6488-6496.
71. Sebastiani F, Barberio C, Casalone E, Cavalieri D, Polsinelli M (2002) Crosses between Saccharomyces cerevisiae and Saccharomyces bayanus generate fertile hybrids. Res Microbiol 153: 53-58.
72. Greig D, Louis EJ, Borts RH, Travisano M (2002) Hybrid speciation in experimental populations of yeast. Science 298: 1773-1775.
73. Liti G, Barton DB, Louis EJ (2006) Sequence diversity, reproductive isolation and species concepts in Saccharomyces. Genetics 174: 839-850.
74. Wei W, McCusker JH, Hyman RW, Jones T, Ning Y, et al. (2007) Genome sequencing and comparative analysis of Saccharomyces cerevisiae strain YJM789. Proc Natl Acad Sci U S A 104: 12825-12830.
75. Doniger SW, Kim HS, Swain D, Corcuera D, Williams M, et al. (2008) A catalog of neutral and deleterious polymorphism in yeast. PLoS Genet 4: e1000183.
76. Esberg A, Muller LA, McCusker JH (2011) Genomic structure of and genome-wide recombination in the Saccharomyces cerevisiae S288C progenitor isolate EM93. PLoS One 6: e25211.
77. Bensasson D, Zarowiecki M, Burt A, Koufopanou V (2008) Rapid evolution of yeast centromeres in the absence of drive. Genetics 178: 2161-2167.

78. Bensasson D (2011) Evidence for a high mutation rate at rapidly evolving yeast centromeres. BMC Evol Biol 11: 211.
79. Boeke JD, Sandmeyer SB (1991) Yeast transposable elements. In: Broach JR, Jones EW, Prongle J, editors. The Molecular and Cdllular Biology of the Yeast Saccharomyces cerevisae. Cold Spring Harbor, N.Y.: Cold Spring Harbor Press. pp. 193-261.
80. Servant G, Pennetier C, Lesage P (2008) Remodeling yeast gene transcription by activating the Ty1 long terminal repeat retrotransposon under severe adenine deficiency. Mol Cell Biol 28: 5543-5554.
81. Knight SA, Labbe S, Kwon LF, Kosman DJ, Thiele DJ (1996) A widespread transposable element masks expression of a yeast copper transport gene. Genes Dev 10: 1917-1929.
82. Gaisne M, Becam AM, Verdiere J, Herbert CJ (1999) A 'natural' mutation in Saccharomyces cerevisiae strains derived from S288c affects the complex regulatory gene HAP1 (CYP1). Curr Genet 36: 195-200.
83. Jordan IK, McDonald JF (1999) The role of interelement selection in Saccharomyces cerevisiae Ty element evolution. J Mol Evol 49: 352-357.
84. Bergman CM, Bensasson D (2007) Recent LTR retrotransposon insertion constrasts with waves of non-LTR insertion since speciation in *Drosophila melanogaster*. Proc Natl Acad Sci U S A 104: 11340-11345.

# Figure Legends

**Figure 1. Example re-annotation of *Ty* elements in the *S. cerevisiae* genome.** Re-annotation of *Ty* elements using RepeatMasker and REannotate is shown in black above, and the original annotation of *Ty* elements and tRNA genes are shown in blue below. The red arrow indicates a newly annotated *Ty1* fragment 5' to a full-length *Ty2* element that joins with a pre-existing *Ty1* fragment 3' to the *Ty2* element, revealing a new TE nesting event.

**Figure 2. Maximum likelihood phylogeny of LTR sequences from the *Ty3/Ty3p* super-family within *Saccharomyces sensu stricto*.** The phylogeny was created from 983 aligned nucleotide positions. The branches are colour-coded based on the species labelled on the phylogeny. Individual sequence names have been omitted. Branches are drawn to scale and the bar represents the number of nucleotide substitutions per site. A "*" denotes branches with both mlBP $\geq$ 70% and biPP $\geq$ 0.95 support.

**Figure 3. Maximum likelihood phylogeny of *Ty1* LTR sequences from the reference genome of *S. cerevisiae*.** The phylogeny was created from 731 aligned nucleotide positions. LTRs in red are from partial elements and are no longer capable of autonomous transposition; LTRs in blue are from full-length elements and denoted as being from the 5' or 3' end of the element. Canonical, *Ty1'* and hybrid *Ty1/2* elements are boxed in green, red and blue, respectively. Clades composed entirely of inactive elements have been collapsed and labelled "Long branch solo LTRs", while recently active solo-LTRs are individually named. Branches are drawn to scale and the bar represents the number of nucleotide substitutions per site. A "/" denotes arbitrarily shortened branches and a "*" denotes branches with both mlBP $\geq$ 70% and biPP $\geq$ 0.95 support.

**Figure 4. Maximum likelihood phylogeny of *Ty2* LTR sequences from the reference genome of *S. cerevisiae*.** The phylogeny was created from 362 aligned nucleotide positions. Colour codes and nomenclature for LTRs are the same as in Figure 3, with the exception that no clades are composed entirely of inactive elements and so none have been collapsed.

**Figure 5. Maximum likelihood phylogeny of *Ty3* and *Ty3p* LTR sequences from the reference genome of *S. cerevisiae*.** The phylogeny was created from 625 aligned nucleotide positions. Green boxes highlight long-branch clades whose positions among the short-branch, younger elements are probably phylogenetic artefacts since they found in different positions in the Bayesian analysis. Formatting of the tree is otherwise the same as in Figure 4.

**Figure 6. Maximum likelihood phylogeny of *Ty4* LTR sequences from the reference genome of *S. cerevisiae*.** The phylogeny was created from 450 aligned nucleotide positions. Formatting of the tree and labels is the same as in Figure 4.

**Figure 7. Maximum likelihood phylogeny of *Ty5* LTR sequences from the reference genome of *S. cerevisiae*.** The phylogeny was created from 268 aligned nucleotide positions. Formatting of the tree and labels is the same as in Figure 4.

**Figure 8. Maximum likelihood phylogeny of the *Ty1/Ty2* super-family within *Saccharomyces sensu stricto*.** The phylogeny was created from 294 aligned LTR nucleotide positions. The branches have been colour coded based on the species from the LTRs were sequenced; species clades are labelled on the phylogeny. Individual sequence names have been omitted. Formatting of the tree and labels is the same as in Figure 2.

# Tables

**Table 1. Copy numbers and fixation status of *Ty* families within the *S. cerevisiae* genome.**

| Family | Total | Full-length | Truncated | Solo LTR | Polymorphic | Fixed | Percentage Fixed |
|---|---|---|---|---|---|---|---|
| *Ty1* | 313 (217) | 32 (32) | 2 (0) | 279 (185) | 48 (4) | 265 | 84.7 |
| *Ty2* | 46 (34) | 13 (13) | 2 (0) | 31 (21) | 35 (1) | 11 | 23.9 |
| *Ty3* | 45 (41) | 2 (2) | 0 (0) | 43 (39) | 22 (1) | 22 | 48.9 |
| *Ty3p* | 15 (0) | 0 (0) | 0 (0) | 15 (0) | 1 (0) | 14 | 93.3 |
| *Ty4* | 49 (32) | 3 (3) | 1 (0) | 45 (29) | 18 (0) | 31 | 63.3 |
| *Ty5* | 15 (7) | 1 (1) | 0 (0) | 14 (6) | 2 (0) | 13 | 86.7 |
| Total | 483 (331) | 51 (51) | 5 (0) | 427 (280) | 127 (6) | 356 | 73.7 |

Full-length elements are defined as those that possess both LTRs; truncated elements have lost one LTR and/or internal DNA through deletion other than LTR-LTR recombination; solo LTRs are generated through intra-element LTR-LTR recombination and can be complete or partial LTR sequences. Numbers in parentheses for the total, full-length, truncated and solo LTR columns are those reported in Kim *et al.* [22]. Fixed elements are those where a particular insertion was present in all scorable SGRP genomes. Numbers in parentheses for the polymorphic column indicate the number of insertions that could only be found in the S288c reference genome.

**Table 2. Patterns of nucleotide diversity among paralogous LTR sequences within *Ty* families in the *S. cerevisiae* genome.**

| | All | | | Inactive | | | Active | | | Mean Terminal Branch Length (subs/site ±S.D.) | |
|---|---|---|---|---|---|---|---|---|---|---|---|
| Family | n | π | θ | n | π | θ | n | π | θ | Tajima's *D* | Fixed | Polymorphic |
| *Ty1* | 313 | 0.300 | 0.308 | 272 | 0.321 | 0.314 | 41 | 0.076 | 0.054 | 1.176 | 0.210±0.151 | 0.020±0.062 |
| *Ty2* | 44 | 0.046 | 0.088 | 9 | 0.119 | 0.142 | 35 | 0.025 | 0.029 | -0.900 | 0.044±0.091 | 0.003±0.011 |
| *Ty3* | 45 | 0.116 | 0.173 | 21 | 0.192 | 0.218 | 24 | 0.019 | 0.035 | -1.878 | 0.212±0.165 | 0.005±0.007 |
| *Ty3p* | 15 | 0.397 | 0.349 | 15 | 0.411 | 0.351 | 0 | - | - | - | 0.374±0.120 | 0.359±0.032 |
| *Ty4* | 49 | 0.304 | 0.251 | 33 | 0.292 | 0.247 | 16 | 0.021 | 0.027 | -1.614 | 0.225±0.107 | 0.004±0.004 |
| *Ty5* | 15 | 0.332 | 0.262 | 15 | 0.322 | 0.256 | 0 | - | - | - | 0.143±0.208 | 0.046±0.064 |

The total number of *Ty2* insertions with LTRs here differs from total number of *Ty2* insertions in Table 1 since two truncated insertions lack LTRs. These are indicated by "n.a." in File S2.

# Supporting Information

**File S1.**

**Format:** .txt

**Title:** Database of *Ty* element reference sequences used for re-annotation of the yeast genome.

**Caption:** Fasta file of LTR and internal sequences from *Ty1-Ty5*, with the inclusion of *Ty3p* from *S. paradoxus*.

**File S2.**

**Format:** .txt

**Title:** Summary of the annotated *Ty* elements in the June 2008 version of *S. cerevisiae* genome from the UCSC Genome Database (sacCer2).

**Caption:** Shown are the REANNOTATE identifier, genomic coordinates of the complete span of all fragments from an individual element, strand, name(s) of corresponding *Ty* elements from the SGD sacCer2 annotation, structural classification, activity status as determined by phylogenetic analysis, estimated allele frequency, and polymorphism/fixation status based on SGRP alignments. More specifically the status of *Ty* insertions in each SGRP alignment is classified as polymorphic or fixed, but *Ty* elements that are present only in the reference strain are labelled as S288c in this table and classified as polymorphic in subsequent analyses. Presence (1), absence (0) or missing data (n.a.) status is shown for individuals SGRP strains in columns following the "sgrp alignment status" column. REANNOTATE identifiers are comprised of the chromosome name, the insertion type, a numerical identifier and *Ty* family name. Insertion type can be full-length (i or u), truncated (t) or solo LTR (s or st). Insertions from the *Ty3p* family are labelled as TY3_1p according to RepBase nomeclature.

**File S3.**

**Format:** .txt

**Title:** GFF file of Repeatmasker/REANNOTATE *Ty* fragments in the June 2008 version of *S. cerevisiae* genome from the UCSC Genome Database (sacCer2).

**Caption:** Individual fragments from the same element are given the same name in the ID column. The span of the union of fragments is the same as the range of coordinates given in File S2.

**File S4.**

**Format:** .txt

**Title:** Nexus file of aligned LTR sequences used for phylogenetic analysis of *Ty3*/*Ty3p* superfamily.

**Caption:** Alignments of paralogous LTR sequence from the *Ty3/Ty3p* superfamily used to create the phylogeny in Figure 2. GenBank Accession Numbers are shown for sequences from species other than *S. cerevisiae*.

**File S5.**

**Format:** .txt

**Title:** Nexus file of aligned LTR sequences used for phylogenetic analysis of *Ty1/Ty2* superfamily.

**Caption:** Alignments of paralogous LTR sequence from *Ty1/Ty2* superfamily used to create the phylogeny in Figure 8. Positions enclosed in square brackets were excluded from the phylogenetic analyses. GenBank Accession Numbers are shown for sequences from species other than *S. cerevisiae*.

**File S6.**

**Format:** .zip

**Title:** ZIP archive of LTR alignments used for molecular evolutionary analyses.

**Caption:** Alignments of paralogous LTR sequence from all, active and inactive elements are provided for *Ty1*, *Ty2*, *Ty3+Ty3p*, *Ty4* and *Ty5*, respectively. For full-length element only the 5' LTR was included. Also included are alignments of the *Ty3/Ty3p* superfamily and the *Ty1/Ty2* superfamily, respectively. Positions enclosed in square brackets were excluded from the phylogenetic analyses. GenBank Accession Numbers are shown for sequences from species other than *S. cerevisiae*.

**File S7.**

**Format:** .zip

**Title:** ZIP archive of SGRP strain alignments used for population genomic analyses.

**Caption:** Alignments of the orthologous genomic region in each SGRP strain corresponding to each *Ty* insertion in our re-annotation of the *S. cerevisiae* reference genome. N's represent low quality (<Q40 phred score) or missing data, primarily due to the low-coverage sequence of many SGRP strains. The DNA sequence of sacCer2 is included as "REF" and each alignment includes the TE locus and 200bp of DNA sequence on either side of the *Ty* insertion site. A dummy sequence shows the position of each TE in the alignment with "T"s.

**File S8.**

**Format:** .pdf

**Title:** Visualization of recombination breakpoints in *Ty1*, *Ty1/2* and *Ty2* complete LTR sequences.

**Caption:** The two approximate recombination breakpoints are highlighted with black lines. Asterisks highlight columns in the alignment where the base is conserved across all sequences.

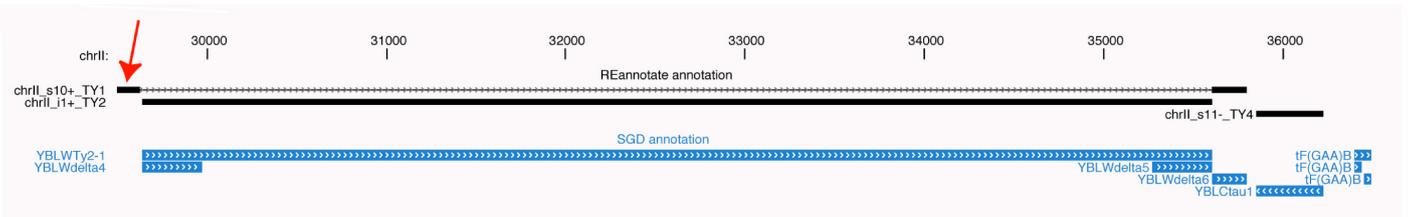

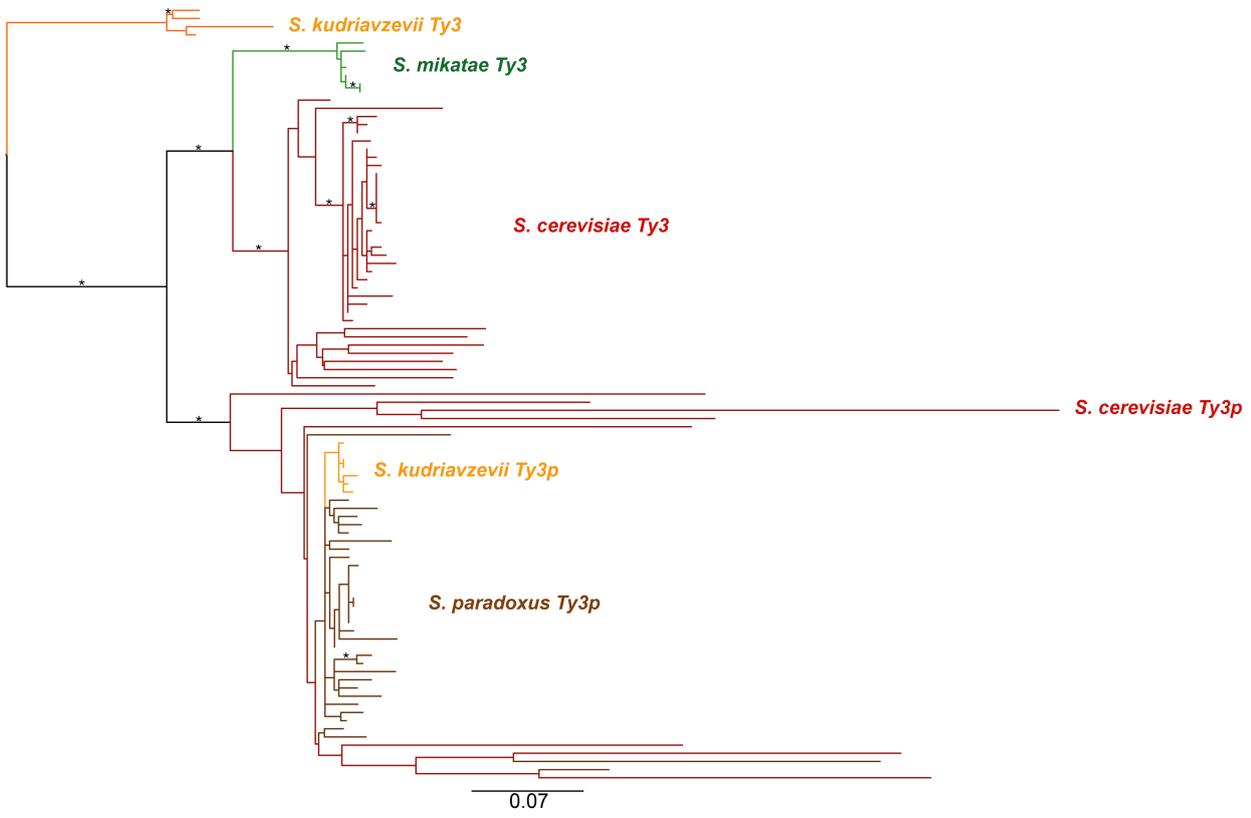

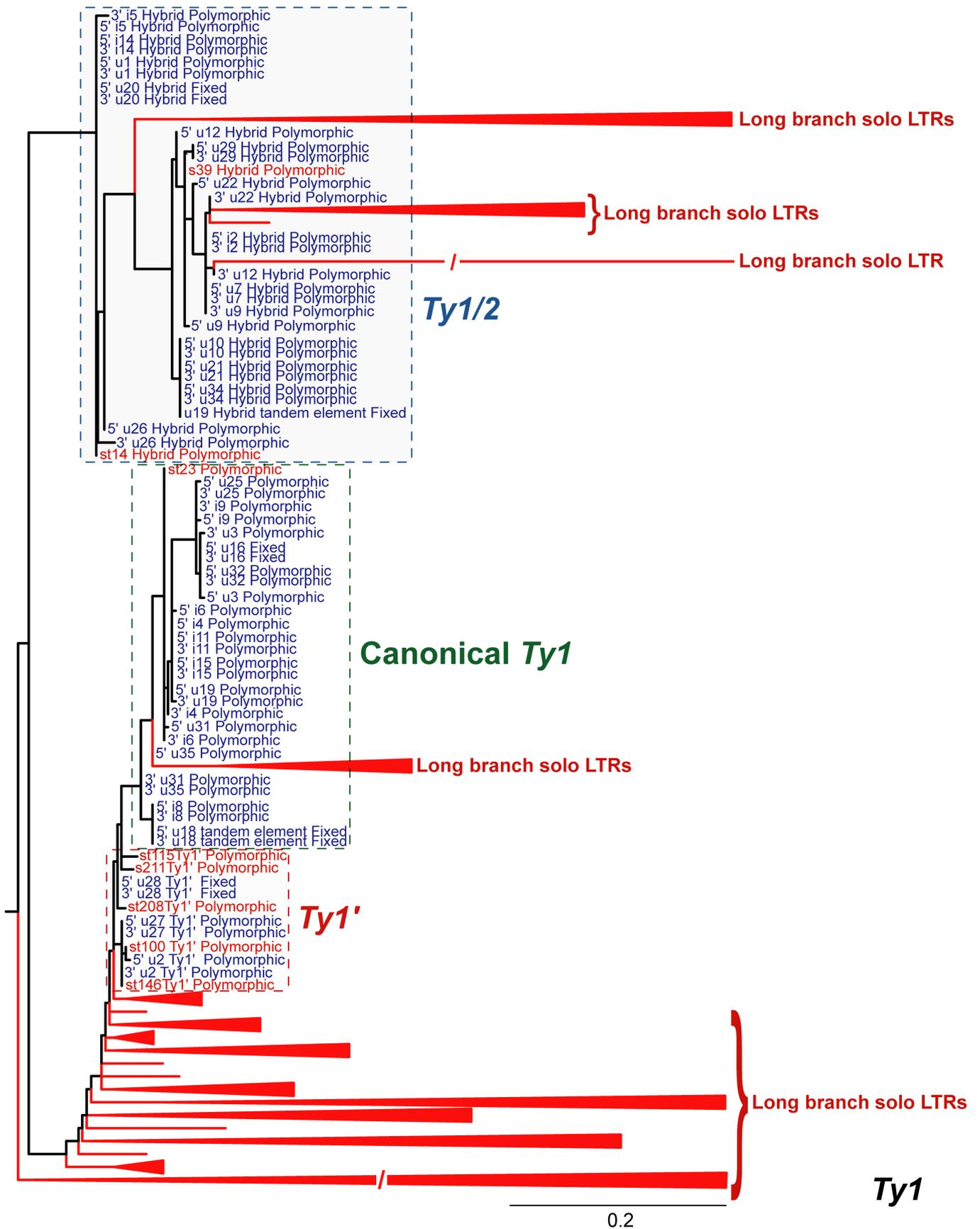

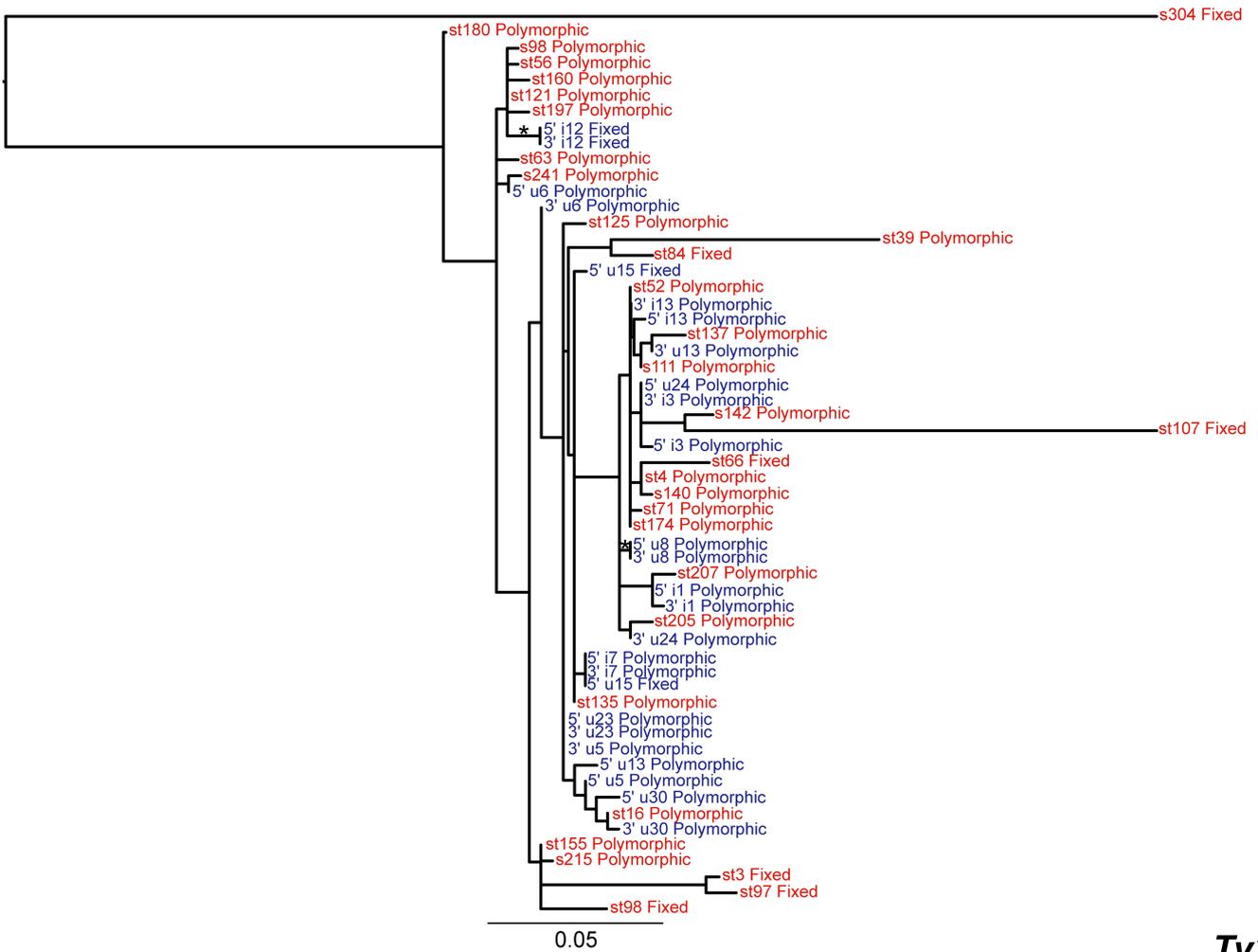

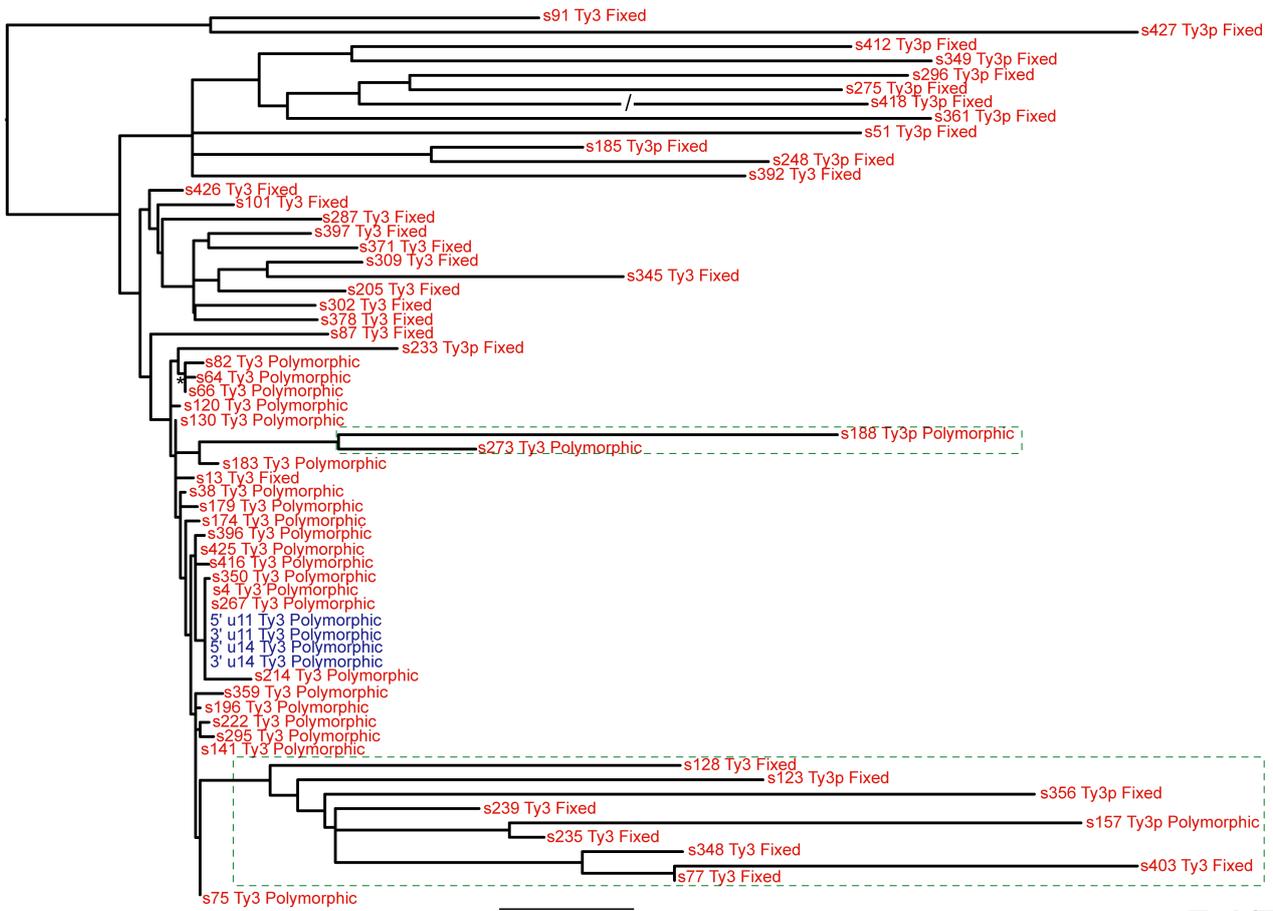

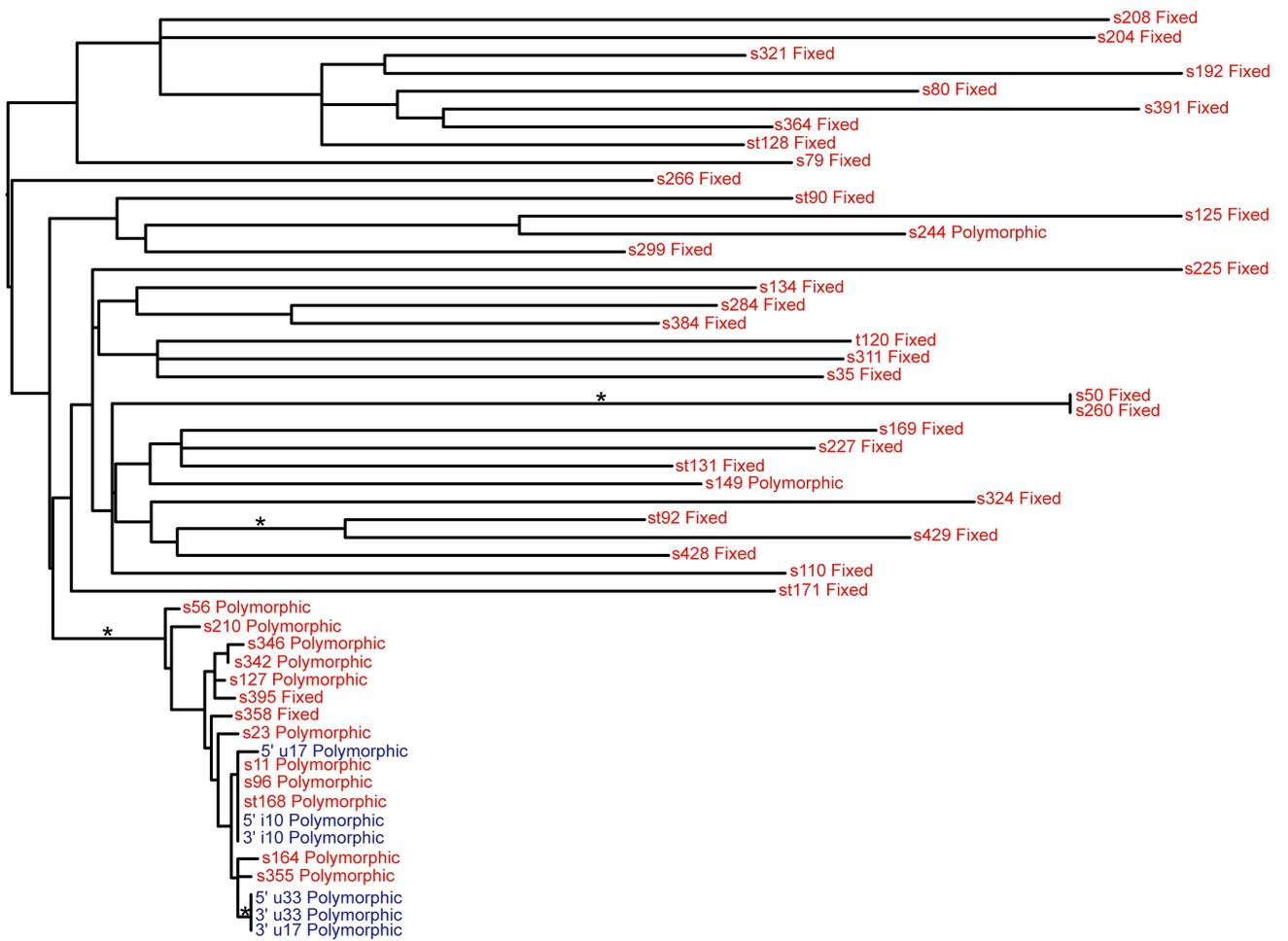

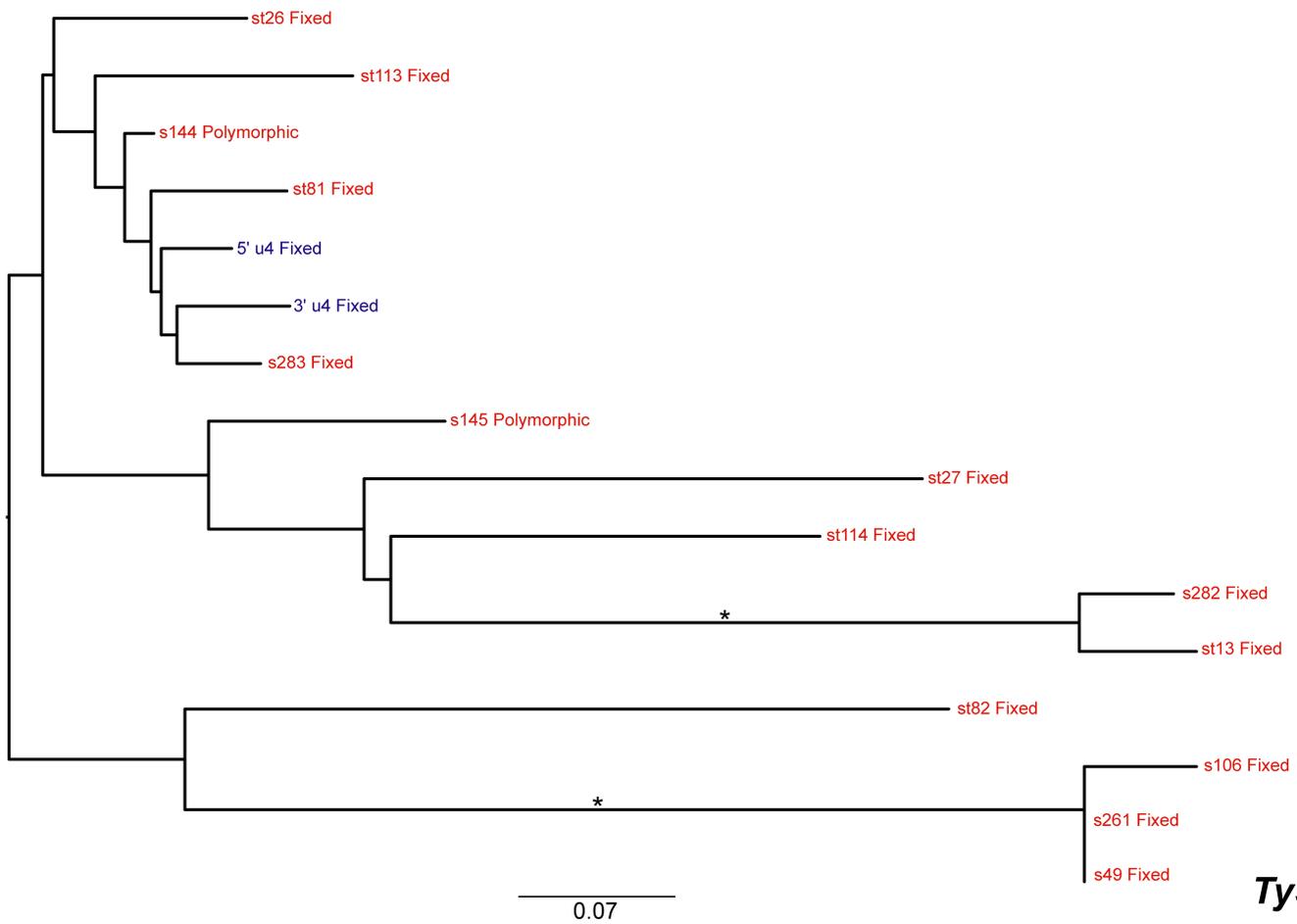

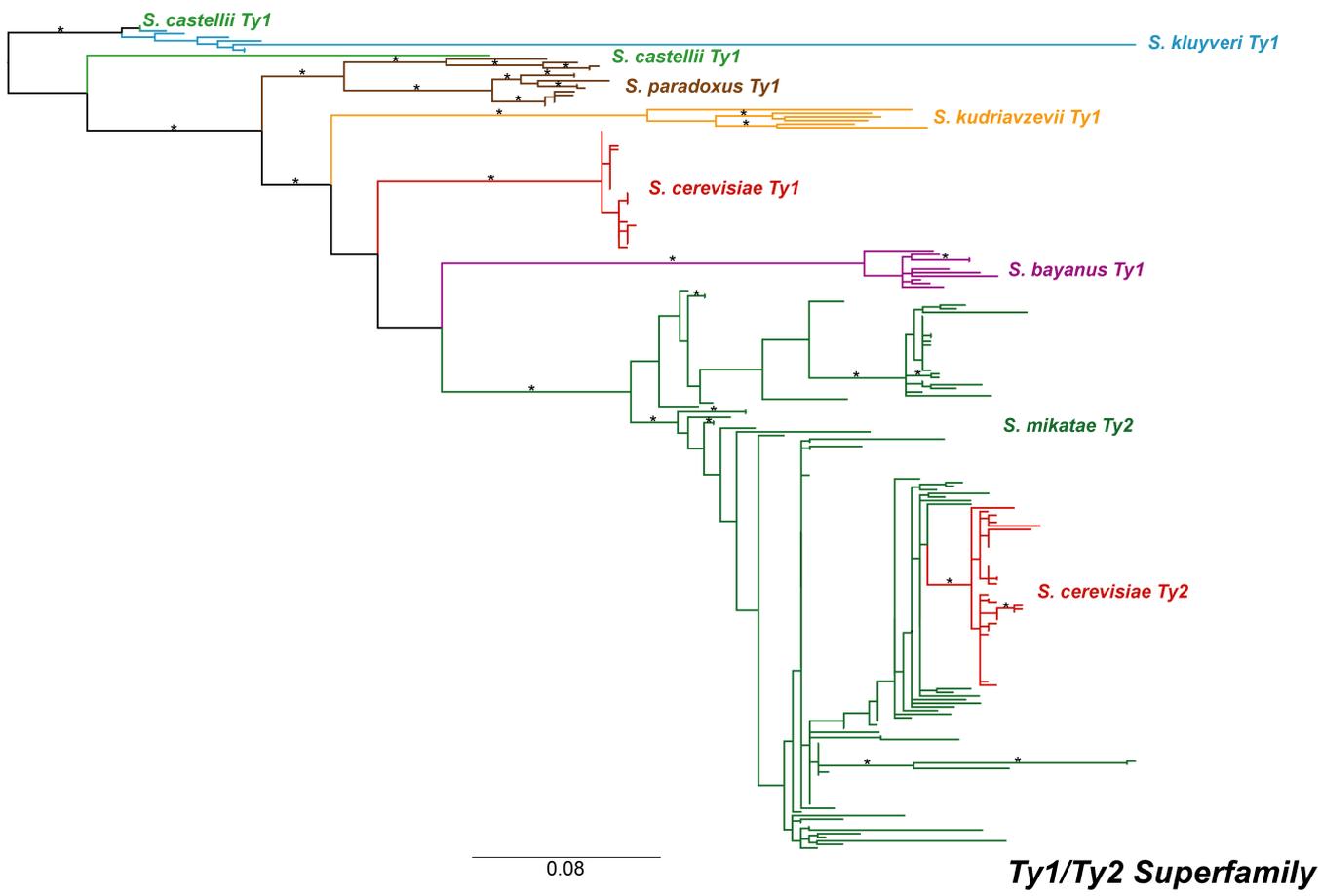